\documentclass[11pt,twoside]{article}
\usepackage{amssymb}   

\def\tr{{\hbox{\rm Tr}}}
\def\ex{{\hbox{\rm e}}}
\def\k{{\cal K}}

\newcommand{\beq}{\begin{equation}}
\newcommand{\eeq}{\end{equation}}
\newcommand{\bear}{\begin{eqnarray}}
\newcommand{\eear}{\end{eqnarray}}

\newcommand{\CS}{{\scriptstyle {\rm CS}}}
\newcommand{\CSs}{{\scriptscriptstyle {\rm CS}}}
\newcommand{\ie}{{\it i.e.}}

\linethickness{.05 pt}
\newsavebox{\unoj}
\savebox{\unoj}(10,3){
\put(2,5) {\circle{10}}
\put(2,0){\line(0,0){10}
\put(0,4){${\scriptscriptstyle j}$}} }
\newcommand{\cunoj}{\mbox{\usebox{\unoj} \hskip 15pt  } } 

\newsavebox{\unodos}
\savebox{\unodos}(10,3){
\put(2,5) {\circle{10}}
\put(2,0){\line(0,0){10}
\put(0,4){${\scriptscriptstyle 2}$}} }
\newcommand{\cunodos}{\mbox{\usebox{\unodos} \hskip 15pt  } } 

\newsavebox{\unotres}
\savebox{\unotres}(10,3){
\put(2,5) {\circle{10}}
\put(2,0){\line(0,0){10}
\put(0,4){${\scriptscriptstyle 3}$}} }
\newcommand{\cunotres}{\mbox{\usebox{\unotres} \hskip 15pt  } } 

\newsavebox{\unocuatro}
\savebox{\unocuatro}(10,3){
\put(2,5) {\circle{10}}
\put(2,0){\line(0,0){10}
\put(0,4){${\scriptscriptstyle 4}$}} }
\newcommand{\cunocuatro}{\mbox{\usebox{\unocuatro} \hskip 15pt  } } 

\newsavebox{\dosi}
\savebox{\dosi}(10,3){
\put(2,5) {\circle{10}}
\put(0,0.3){\line(0,1){9.2}}
\put(4,0.3){\line(0,1){9.2} } }

\newsavebox{\dosii}
\savebox{\dosii}(10,3){
\put(2,5) {\circle{10}}
\put(-3,5){\line(3,0){10}}
\put(2,0){\line(0,5){10} } }
\newcommand{\cdosii}{\mbox{\usebox{\dosii} \hskip 15pt  } } 

\newsavebox{\dosiidosuno}
\savebox{\dosiidosuno}(10,3){
\put(2,5) {\circle{10}}
\put(-3,5){\line(3,0){10}}
\put(2,0){\line(0,5){10}
\put(0,7){${\scriptscriptstyle 2}$} } }
\newcommand{\cdosiidosuno}{\mbox{\usebox{\dosiidosuno} \hskip 15pt  } } 

\newsavebox{\dosiidosdos}
\savebox{\dosiidosdos}(10,3){
\put(2,5) {\circle{10}}
\put(-3,5){\line(3,0){10}}
\put(2,0){\line(0,5){10}
\put(0,7){${\scriptscriptstyle 2}$}
\put(-3,5){${\scriptscriptstyle 2}$} } }
\newcommand{\cdosiidosdos}{\mbox{\usebox{\dosiidosdos} \hskip 15pt  } } 

\newsavebox{\dosiitresuno}
\savebox{\dosiitresuno}(10,3){
\put(2,5) {\circle{10}}
\put(-3,5){\line(3,0){10}}
\put(2,0){\line(0,5){10}
\put(0,7){${\scriptscriptstyle 3}$} } }
\newcommand{\cdosiitresuno}{\mbox{\usebox{\dosiitresuno} \hskip 15pt  } } 

\newsavebox{\tresiii}
\savebox{\tresiii}(10,3){
\put(2,5) {\circle{10}}
\put(0,0.3){\line(0,1){9.2}}
\put(-3,5){\line(3,0){10}}
\put(4,0.4){\line(0,1){9.1} } }
\newcommand{\ctresiii}{\mbox{\usebox{\tresiii} \hskip 15pt  }
} 

\newsavebox{\tresiv}
\savebox{\tresiv}(10,3){
\put(2,5) {\circle{10}}
\put(-0.25,0.5){\line(1,2){4.45}}
\put(-3,5){\line(3,0){10}}
\put(4.25,0.6){\line(-1,2){4.45} } }
\newcommand{\ctresiv}{\mbox{\usebox{\tresiv} \hskip 15pt  }
} 

\newsavebox{\tresivdos}
\savebox{\tresivdos}(10,3){
\put(2,5) {\circle{10}}
\put(-0.25,0.5){\line(1,2){4.45}}
\put(-3,5){\line(3,0){10}}
\put(4.25,0.6){\line(-1,2){4.45}
\put(-4,4.5){${\scriptscriptstyle 2}$} } }
\newcommand{\ctresivdos}{\mbox{\usebox{\tresivdos} \hskip 15pt  }
} 

\newsavebox{\tresiiidos}
\savebox{\tresiiidos}(10,3){
\put(2,5) {\circle{10}}
\put(0,0.3){\line(0,1){9.2}}
\put(-3,5){\line(3,0){10}}
\put(4,0.4){\line(0,1){9.1}
\put(-4,6){${\scriptscriptstyle 2}$} } }
\newcommand{\ctresiiidos}{\mbox{\usebox{\tresiiidos} \hskip 15pt  }
} 

\newsavebox{\tresiiidosbis}
\savebox{\tresiiidosbis}(10,3){
\put(2,5) {\circle{10}}
\put(0,0.3){\line(0,1){9.2}}
\put(-3,5){\line(3,0){10}}
\put(4,0.4){\line(0,1){9.1}
\put(-2.5,5){${\scriptscriptstyle 2}$} } }
\newcommand{\ctresiiidosbis}{\mbox{\usebox{\tresiiidosbis} \hskip 15pt  }
} 

\newsavebox{\cuavi}
\savebox{\cuavi}(10,3){
\put(2,5) {\circle{10}}
\put(-1,0.8){\line(0,1){8.2}}
\put(2,0){\line(0,5){10} }
\put(5,1){\line(0,1){7.8} }
\put(-3,5){\line(3,0){10} }}
\newcommand{\ccuavi}{\mbox{\usebox{\cuavi} \hskip 15pt  } } 

\newsavebox{\cuavii}
\savebox{\cuavii}(10,3){
\put(2,5) {\circle{10}}
\put(-1,0.8){\line(0,1){8.1}}
\put(-2.7,3.1){\line(5,6){5.6} }
\put(6.5,3.1){\line(-5,6){5.6} }
\put(5,1){\line(0,1){7.8} } }
\newcommand{\ccuavii}{\mbox{\usebox{\cuavii} \hskip 15pt  }
} 

\newsavebox{\cuaviii}
\savebox{\cuaviii}(10,3){
\put(2,5) {\circle{10}}
\put(-1,0.8){\line(0,1){8.2}}
\put(2,0){\line(1,3){2.99} }
\put(5,1){\line(-1,3){3} }
\put(-3,5){\line(3,0){10} }}
\newcommand{\ccuaviii}{\mbox{\usebox{\cuaviii} \hskip 15pt  }
}

\newsavebox{\cuaix}
\savebox{\cuaix}(10,3){
\put(2,5) {\circle{10}}
\put(0,0.3){\line(0,1){9.2}}
\put(4,0.3){\line(0,1){9.2} }
\put(-2.7,3){\line(3,0){9.2}}
\put(-2.7,7){\line(3,0){9.2}} }
\newcommand{\ccuaix}{\mbox{\usebox{\cuaix} \hskip 15pt  }
} 

\newsavebox{\cuax}
\savebox{\cuax}(10,3){
\put(2,5) {\circle{10}}
\put(-0.3,0.5){\line(1,2){4.5}}
\put(4.3,0.6){\line(-1,2){4.45} }
\put(-2.7,3){\line(3,0){9.2}}
\put(-2.7,7){\line(3,0){9.2}} }
\newcommand{\ccuax}{\mbox{\usebox{\cuax} \hskip 15pt  }
} 

\newsavebox{\cuaxi}
\savebox{\cuaxi}(10,3){
\put(2,5) {\circle{10}}
\put(-3,5){\line(3,0){10}}
\put(2,0){\line(0,5){10} }
\put(-1.6,1.4){\line(1,1){7.1} }
\put(5.5,1.5){\line(-1,1){7.05} }}
\newcommand{\ccuaxi}{\mbox{\usebox{\cuaxi} \hskip 15pt  } } 

\newsavebox{\extra}
\savebox{\extra}(10,3){
\put(2,5) {\circle{10}}
\put(0,0.3){\line(0,1){9.2}}
\put(4,0.4){\line(0,1){9.1} } }
 
\begin{document}

\pagestyle{myheadings}
\markboth
{{\sc Knot Theory from a Chern-Simons...
}}   {{\sc J. M. F. Labastida  }}        
\vspace{1cm}

%
%

\thispagestyle{empty}

\begin{flushright}
USC-FT-3/00 \\
hep-th/0002221 
\end{flushright}
\vglue 1cm

\begin{center}

{\Large{\bf{Knot Theory from a Chern-Simons}}} 

\medskip

{\Large{\bf{Gauge Theory Point of View\footnotemark}}}     

\vskip1cm

{\sc J. M. F. Labastida}   
\vskip0.5cm

\footnotetext{Invited lecture delivered at the Workshop on
Geometry and Physics, held at Medina del Campo, Spain,
September 20 -- 22, 1999.}

{\it Departamento de F\'\i sica de Part\'\i
culas \\Universidade de Santiago de Compostela\\
15706 Santiago de Compostela, Spain\\
{\rm e-mail: labasti@fpaxp1.usc.es}}         
\vskip0.3cm
\end{center}
\bigskip

%
%

\begin{abstract}
A brief summary of the development of perturbative 
Chern-Simons gauge theory related to the theory of knots and
links is presented. Emphasis is made on the progress achieved
towards the determination of a general combinatorial
expression for Vassiliev invariants. Its form for all the
invariants  up to order four is reviewed, and a table of their
values for all prime knots with ten crossings is presented.
\end{abstract}

%
%

\section{Introduction}

Chern-Simons gauge theory \cite{csgt,primeros} provides an excellent framework
to study knot and link invariants. This framework has shown
to be very useful in both, the study of polynomial invariants as the
Jones polynomial \cite{jones} and its generalizations, and the study of
Vassiliev invariants or numerical invariants of finite type.
Non-perturbative aspects of the theory play a fundamental role in
the first context while perturbative ones are the main tool
in the second. In this paper I will present a brief summary of
the results on Vassiliev \cite{vass,birrev,barnatan,bilin, singular}
invariants achieved from perturbative Chern-Simons gauge theory in the last
few years. 

An important line of investigation in the context of Vassiliev
invariants is the search of a universal combinatorial
expression. Different approaches \cite{lannes,arrgpo,virgpo,temporal} have
been carried out. In the framework of perturbative Chern-Simons gauge theory
explicit combinatorial formulae for all the primitive invariants up to order
four have been obtained \cite{temporal}. The context based on this approach
seems rather promising to obtain a general combinatorial expression.

As any other gauge theory, Chern-Simons gauge theory can be
analyzed for different gauge fixings. Covariant gauges lead to
complicated integrals expressions for Vassiliev invariants
\cite{gmm,natan,alla,alemanes,alts,bt,dylan}. Simpler integral expressions are
obtained in non-covariant gauges of the light-cone type
\cite{vande,kont,cata,lcone,kaucone}. However, for a non-covariant gauge
fixing of the temporal type one obtains combinatorial expressions
\cite{temporal,faroknots}. It is in the last situation in the one that all
intermediate integrals can be done so one ends with  combinatorial
expressions where only the information contained at the crossings is relevant.

Combinatorial expressions are much preferred to compute
invariants and to study their properties. It turns out that the
resulting combinatorial expressions can be rewritten in terms
of Gauss diagrams. These facts notably simplify their
explicit formulae. In this brief presentation I will collect all
the expressions based on Gauss diagramas for all the primitive
Vassiliev invariants up to order four.

An extended version of this presentation can be found in ref.
\cite{baires}. In the present paper, however, I include a table that was not
presented in ref. \cite{baires} because the corresponding computations
were not carried out then. The appendix contains a table with all
the primitive Vassiliev invariants up to order four for all prime
knots with ten crossings, as computed from the combinatorial
expressions provided by Chern-Simons gauge theory.

The paper is organized as follows. In sect. 2 I briefly
describe the quantization procedure in the temporal gauge and
comment on the essential ingredients which lead to the
combinatorial expressions. In sect. 3 the resulting
combinatorial expressions are presented in terms of Gauss
diagrams. Finally, in sect. 4 aspects of
future research directions are described.

\section{Perturbative Chern-Simons gauge theory in the temporal
gauge}

I will begin reviewing the basic elements of Chern-Simons gauge
theory. This theory is a quantum field theory whose action is
based on the Chern-Simons form associated to a non-abelian
gauge group.  The fundamental data in Chern-Simons gauge theory
are the following: a smooth three-manifold $M$ which  will be
taken to be compact, a gauge group $G$, which will be taken
semi-simple and compact, and an integer parameter $k$.  The
action of the theory is the integral of the Chern-Simons form
associated to a gauge connection $A$ corresponding to a gauge
group $G$:
\begin{equation}
S_{\CS} (A) ={k\over 4\pi} \int_M \tr (A\wedge d A + \frac{2}{3} A\wedge
A\wedge A).
\label{valery}
\end{equation}
 The exponential of $i$ times
this action is invariant under gauge transformations for integer
$k$.

The metric-independence of the action (\ref{valery}) 
implies that the resulting quantum field theory is topological.
Appropriate observables lead to vacuum expectation values (vevs)
which correspond to topological invariants. A particularly
important set of observables is constituted by Wilson loops.
They  correspond to the holonomy of the gauge connection $A$
along a loop. Given a representation
$R$ of the gauge group $G$ and a 1-cycle
$\gamma$ on $M$, it is defined as:
\begin{equation}
W_\gamma^R (A) =\tr_R \big( {\hbox{\rm Hol}}_\gamma (A) \big) =
\tr_R {\hbox{\rm P}} \exp \int_\gamma A.
\label{rosalia}
\end{equation}
Products of these operators are the natural candidates to obtain
topological invariants after computing their vev. These
vevs are formally written as: 
\begin{equation}
\langle W_{\gamma_1}^{R_1} W_{\gamma_2}^{R_2} \cdots
W_{\gamma_n}^{R_n} \rangle 
        \nonumber \\
= \int [DA] W_{\gamma_1}^{R_1}(A) W_{\gamma_2}^{R_2}(A) \cdots
W_{\gamma_n}^{R_n}(A)   \ex^{i S_{\CSs} (A) },
\label{encarna}
\end{equation}
where $\gamma_1$, $\gamma_2$,$\dots$,$\gamma_n$ are 1-cycles on $M$ and
$R_1$,
$R_2$ and $R_n$ are representations of $G$. In (\ref{encarna}),
the quantity $[DA]$ denotes the functional integral measure. The
functional integral in  (\ref{encarna}) is not well defined.
A variety of methods have been proposed to go
around this problem and provide some meaning to the right hand
side of (\ref{encarna}). These methods fall into two categories,
perturbative and non-perturbative ones.   Witten, in his
pioneer work 1988 \cite{csgt}, showed, using non-perturbative
methods, that when one considers non-intersecting cycles
$\gamma_1$, $\gamma_2$,$\dots$,$\gamma_n$ without
self-intersections, the vevs (\ref{encarna}) lead to
polynomial invariants as the Jones polynomial and
its generalizations. 

 The construction of
the perturbative series expansion of the vev of an operator when
dealing with a gauge theory starts with a gauge
fixing. The first analysis of Chern-Simons perturbation
theory were made in the covariant Landau gauge \cite{gmm}. Subsequent
studies \cite{natan,alla,alts} in this gauge led to a framework
linked to the theory of Vassiliev invariants, which culminated with the
configuration space integral approach \cite{bt,dylan}. 

Non-covariant gauges have been also studied in the context of
Chern-Simons perturbation theory. The
perturbative series which results in the non-covariant
light-cone gauge \cite{lcone} leads to the Kontsevich integral \cite{kont}. 
Vevs of gauge invariant operators are independent of the gauge chosen
and therefore the expressions obtained in the light-cone gauge
should be equivalent to the ones obtained in any other gauge.
Thus, from a quantum field theory point of view, the
configuration space integral which appear in the covariant
Landau gauge leads to the same quantities as the Kontsevich
integral. The non-covariant temporal gauge leads to alternative
expressions for the vevs of gauge-invariant operators which
turn out to be combinatorial \cite{temporal,faroknots}. Again, gauge
invariance implies that the resulting quantities must be the same as the ones
in the configuration space approach or in the Kontsevich integral.
In the rest of this section I will review the salient features
of the analysis of the perturbative series expansion of the
vacuum expectation value of a Wilson loop in the temporal
gauge.

The gauge-fixing condition in the temporal gauge takes the form
\begin{equation} 
n^\mu A_\mu =0, 
\label{light} 
\end{equation}  
where $n$ is the unit vector $n^\mu=(1,0,0)$. In this gauge the
propagator becomes:
\begin{equation}
\delta_{ab} {\lambda \over (np)^2}(p_\mu p_\nu - {i\over \lambda} (np)
\epsilon_{\mu\nu\rho} n^\rho )  \rightarrow
-i \epsilon_{\mu\nu\rho} {n^\rho \over n p} \delta_{ab},
\label{nprop}
\end{equation}
 where the limit $\lambda \rightarrow 0$ has been taken. To
construct the perturbative series expansion of the vev of a
Wilson loop, one needs the Fourier transform  of (\ref{nprop}).
 In the temporal gauge the momentum-space
integral that must  be carried out has the form: 
\begin{equation} 
\Delta(x_0,x_1,x_2) = \int_M {{\rm d} ^3p\over (2\pi)^3}
{\ex^{i(p^0 x_0 +  p^1 x_1 + p^2 x_2)} \over p^0}. 
\label{espacio} 
\end{equation}  
This integral is ill-defined due to the pole at $p^0=0$. To
make sense of it a prescription has to be given to circumvent
the pole. A precise prescription will not be taken. Instead,
 a rather general form will be used \cite{temporal},
\begin{equation} 
\Delta(x_0,x_1,x_2)= {i\over 2} {\hbox{\rm
sign}}(x_0) \delta(x_1)\delta(x_2) +f(x_1,x_2), 
\label{solu} 
\end{equation}
where
$f(x_1,x_2)$ is a prescription-dependent distribution. 
The important
consequence  of the result (\ref{solu}) is that the  
dependence of
$\Delta(x_0,x_1,x_2)$ on
$x_0$ has to be of the form  ${\hbox{\rm sign}}(x_0) 
\delta(x_1)\delta(x_2)$.  This observation will be crucial in 
our analysis. The
propagator (\ref{solu}) will allow us to introduce the notion of
kernel of a Vassiliev invariant and to design a procedure to
compute combinatorial expressions for these invariants.

Given a knot $K$ and one of its regular knot
projections, ${\cal K}$, on the $x_1,x_2$-plane which is a
Morse knot in the $x_1$ and $x_2$ directions, one has to deal
with the following perturbative series expansion for the vacuum
expectation value of the corresponding Wilson loop \cite{temporal}:
\beq \langle W({K},G)\rangle = \langle
W({\k},G)\rangle_{{\rm temp}}
\times  \langle W(U,G)\rangle^{b(\k)}, 
\label{global} 
\eeq
being,
\beq
{1 \over d} \langle W({K},G)\rangle = 1 + 
\sum_{i=1}^{\infty} v_i(K) x^i,
\label{expansiona}
\eeq
and,
\beq
{1 \over d} \langle W({\k},G)\rangle_{{\rm temp}} = 
1 + \sum_{i=1}^{\infty} 
\hat v_i(\k)  x^i.
\label{expansionb}
\eeq
In these expressions $x$ denotes the inverse of the
Chern-Simons coupling constant, $x=2\pi i/k$, $G$ the gauge
group, and $d$ the dimension of the representation carried by
the Wilson loop. The function $b({\cal K})$ is the exponent of
the Kontsevich factor, which has been conjectured to be
\cite{temporal},
\beq
b(\k) = {1\over 12} (n_{x_1}+n_{x_2}),
\label{hipotesis}
\eeq
where $n_{x_1}$ and $n_{x_2}$ are the critical points of the
regular projection ${\cal K}$ in both, the $x_1$ and the $x_2$
directions. In (\ref{global}) $U$ denotes the unknot and 
$\langle W({\k},G)\rangle_{{\rm temp}}$ is the vacuum
expectation of the Wilson loop corresponding to the regular
projection ${\cal K}$ as computed perturbatively in the
temporal gauge with the standard Feynman rules of the theory.
Notice that though each of the factors on the right hand side
of (\ref{global}) depends on the regular projection chosen, the
left hand side does not. While the coefficients $v_i(K)$ of the
series (\ref{expansiona}) are Vassiliev invariants the
coefficients $\hat v_i(K)$  of (\ref{expansionb}) are not.  The
latter depend on the regular projection chosen.

\section{Combinatorial expressions in terms of Gauss diagrams}

A universal combinatorial formula for Vassiliev invariants
could be obtained if the coefficients $\hat v_i(K)$  in
(\ref{expansionb}) could be computed with no integrals left.
Unfortunately, this has not been obtained yet to all orders.
Only part of the contributions entering $\hat v_i(K)$ have been
explicitly written to all orders. These are the kernels
introduced in \cite{temporal}. The kernels are quantities which
depend on the knot projection chosen and therefore are not knot
invariants. However, at a given order $i$ a kernel differs from
an invariant of type $i$ by terms that vanish in signed sums of
order $i$. The kernel contains the part of a Vassiliev invariant which is
the last in becoming zero when performing signed sums, in other
words, a kernel vanishes in signed sums of order $i+1$ but does
not in signed sums of order $i$. 
Kernels plus the structure of the perturbative series expansion
seem to contain enough information to reconstruct the full
Vassiliev invariants \cite{temporal}.

The expression for the kernels results after considering only the
simplest part of the propagator of the gauge field in the temporal gauge.
This part involves a double delta function and therefore all the integrals
can be performed. The result is a combinatorial expression in terms of
crossing signatures after distributing propagators among all the
crossings. The general expression can be written in a universal form much
in the spirit of the universal form of the Kontsevich integral \cite{kont}.
Let us consider a knot $K$ with a regular knot projection ${\cal K}$
containing $n$ crossings. Let us choose a base point on ${\cal K}$ and let
us label the $n$ crossings by $1,2,\dots,n$ as one passes for
first time through each of them when traveling along ${\cal K}$ starting
at the base point. The universal expression for the kernel associated to
${\cal K}$ has the form: 
\begin{eqnarray}
&& {\cal N}(\k) = \sum_{k=0}^\infty\Bigg(
\sum_{m=1}^k \sum_{p_1,\dots,p_m =1\atop p_1+\cdots+p_m=k}^k
\sum_{i_1,\dots,i_m=1\atop i_1 < \cdots < i_m}^n
{\epsilon_{i_1}^{p_1} \cdots \epsilon_{i_m}^{p_m} \over
(p_1!\cdots p_m!)^2}  \nonumber \\ &&
\nonumber \\ && 
\hbox{\hskip5cm}
\sum_{\sigma_{1},\dots,\sigma_{m} \atop
\sigma_{1}\in P_1,\dots,\sigma_{m}\in P_m}
{\cal T}(i_1,\sigma_{1};\dots;i_m,\sigma_{m})\Bigg).
\nonumber \\ 
\label{nucleos}
\end{eqnarray}
In this equation $P_m$ denotes the permutation group of $p_m$ elements. The
factors in the innest sum, ${\cal T}(i_1,\sigma_{1};\dots;i_m,\sigma_{m})$,
are group factors which are computed in the following way:
given a set of crossings, $i_1, \dots, i_{m}$, and a set of permutations,
$\sigma_1,\dots,\sigma_m$, with $\sigma_1\in P_1,\dots,\sigma_m\in P_m$,
the corresponding group factor
${\cal T}(i_1,\sigma_{1};\dots;i_m,\sigma_{m})$ is the result of taking a
trace over the  product
of group generators which is obtained after assigning $p_1,\dots,p_m$ group
generators to the crossings $i_1, \dots, i_{m}$ respectively, and placing
each set of group generators in the order which results after traveling
along the knot starting from the base point. The first time that one
encounters a crossing
$i_j$ a product of
$p_j$ group generators is introduced; the  second
time the product is similar, but with the indices rearranged according to 
the permutation $\sigma_j\in P_j$.

The universal formula (\ref{nucleos}) for the kernels can be written in a
more useful way collecting all the coefficients multiplying a given group
factor. The group factors can be labeled by chord diagrams. At  order
$k$ one has a term for each of the inequivalent chord diagrams with $k$
chords. Denoting chord diagrams by $D$, equation (\ref{nucleos}) can be
written as:
\beq 
{\cal N}(\k) = \sum_{D} N_D(\k) D,
\label{masnucleos}
\eeq
where the sum extends to all inequivalent chord diagrams.
The concept of kernel can be
extended to include singular knots by considering signed sums of
(\ref{masnucleos}), or, following
\cite{singular}, introducing vacuum expectation values of the operators for
singular knots. If $\k^j$ denotes a regular projection of a knot $K^j$ with
$j$ simple singular crossings or double points, the corresponding universal
form for the kernel possesses an expansion similar to  (\ref{masnucleos}):
\beq 
{\cal N}(\k^j) = \sum_{D} N_D(\k^j) D.
\label{sinnucleos}
\eeq
The general results about singular knots proved in \cite{singular} lead to
two important features for (\ref{sinnucleos}). On the one hand,
finite type implies that $N_D(\k^j)=0$ for chord diagrams $D$ with more
than $j$ chords. On the other hand, $N_D(\k^j)=2^j \delta_{D,D({\cal
K}^j)}$, where $D({\cal K}^j)$ is the configuration corresponding to
the singular knot projection ${\cal K}^j$. As observed above, kernels
constitute the part of a Vassiliev invariant which survives a maximum
number of signed sums.

   To compute $N_D(\k)$ one needs to introduce first the notion of the set of
labeled chord subdiagrams of a given chord diagram. This set will be denoted
by $S_D$. This set is made out of a selected set of labeled chord diagrams
that will be defined now. 
A {\it labeled chord diagram} of order $p$ is a chord diagram with $p$
chords and a set of positive integers $k_1,k_2,\dots,k_p$, which will be
called labels, such that each chord has one of these integers attached.
The set $S_D$ is made out of labeled chord diagrams which satisfy two
conditions. These conditions are fixed by the form of the series entering
the kernels (\ref{nucleos}).   The elements of $S_D$ will be called labeled
chord subdiagrams of the chord diagram $D$. They are defined as follows.
A {\it labeled chord subdiagram} of a chord diagram $D$ with $k$ chords is a
labeled chord diagram of order $p$ with labels $k_1,k_2,\dots,k_p$, $p\leq
k$, such that the following two conditions are satisfied: {\it a)}
$k_1+k_2+\cdots+k_p=k$; {\it b)} there exist  elements $\sigma_1 \in
P_{k_1},\sigma_2\in P_{k_2},\dots,\sigma_p\in P_{k_p}$ of the permutation
groups $P_{k_1},P_{k_2},\dots,P_{k_p}$ such that, after replacing the $j$-th
chord diagram by $k_j$ chords arranged according to the permutation
$\sigma_j$, for $j=1,\dots,p$, the resulting chord diagram is homeomorphic
to $D$. The number of ways that permutations  
$\sigma_1 \in P_{k_1},\sigma_2\in P_{k_2},\dots,\sigma_p\in P_{k_p}$ can be
chosen is called the multiplicity of the labeled chord subdiagram. The
multiplicity of a given labeled chord subdiagram, $s\in S_D$, will be denoted 
 by
$m_D(s)$.

The chord diagram $D$ itself can be
regarded as a labeled chord subdiagram such that its labels, or positive
integers attached to its chords, are 1. It has multiplicity 1. All
the elements of $S_D$ except
$D$ have a number of chords smaller than the number of chords of $D$. Not
all labeled chord diagrams are subdiagrams of $D$. However, given a labeled
chord diagram with labels  $k_1,k_2,\dots,k_p$ there can be different sets
of permutations leading to $D$. The number of these different sets is the
multiplicity introduced above. The elements of the sets
$S_D$ for all chord diagrams $D$ up to order four
which do not have disconnected subdiagrams are the following:
\begin{equation}
\vbox{\begin{eqnarray*}
\cdosii &\longrightarrow & \hskip0.3cm \cdosii \hbox{\hskip-0.4cm}, 
\cunodos \\
\vbox{\vskip0.9cm}
\ctresiii &\longrightarrow & \hskip0.3cm \ctresiii \hbox{\hskip-0.4cm}, 
\cdosiidosuno
\hbox{\hskip-0.4cm}, 2
\cunotres
\\
\vbox{\vskip0.9cm}
\ctresiv &\longrightarrow & \hskip0.3cm \ctresiv \hbox{\hskip-0.4cm},
 \cdosiidosuno
\hbox{\hskip-0.4cm},  \cunotres
\\
\vbox{\vskip0.9cm}
\ccuavi &\longrightarrow & \hskip0.3cm \ccuavi \hbox{\hskip-0.4cm}, 
\ctresiiidos \hbox{\hskip-0.4cm}, \cdosiitresuno
\hbox{\hskip-0.4cm},  2 \cunocuatro
\\
\vbox{\vskip0.9cm}
\ccuavii &\longrightarrow & \hskip0.3cm \ccuavii \hbox{\hskip-0.4cm}, 
 2 \cunocuatro
\\
\vbox{\vskip0.9cm}
\ccuaviii &\longrightarrow & \hskip0.3cm \ccuaviii \hbox{\hskip-0.4cm}, 
\ctresiiidos \hbox{\hskip-0.4cm}, 2\cdosiitresuno
\hbox{\hskip-0.4cm},  4 \cunocuatro
\\
\vbox{\vskip0.9cm}
\ccuaix &\longrightarrow & \hskip0.3cm \ccuaix \hbox{\hskip-0.4cm}, 
\ctresiiidosbis \hbox{\hskip-0.4cm}, 2\cdosiidosdos
\hbox{\hskip-0.4cm},   \cunocuatro
\\
\vbox{\vskip0.9cm}
\ccuax &\longrightarrow & \hskip0.3cm \ccuax \hbox{\hskip-0.4cm}, 
\ctresivdos \hbox{\hskip-0.4cm}, \ctresiiidosbis
\hbox{\hskip-0.4cm}, 2\cdosiidosdos \hbox{\hskip-0.4cm},
2\cdosiitresuno \hbox{\hskip-0.4cm}, 3\cunocuatro
\\
\vbox{\vskip0.9cm}
\ccuaxi &\longrightarrow & \hskip0.3cm \ccuaxi \hbox{\hskip-0.4cm}, 
\ctresivdos \hbox{\hskip-0.4cm}, \cdosiidosdos \hbox{\hskip-0.4cm},
\cdosiitresuno \hbox{\hskip-0.4cm}, \cunocuatro
\end{eqnarray*}}
\end{equation}
The numbers accompanying each labeled chord subdiagram denote their
multiplicity. When no number is attached to a chord of a labeled chord
diagram it should be understood that the corresponding label is 1.

In order to write the final expression for the kernels the
notion of Gauss diagram must be introduced. Given a regular projection ${\cal
K}$ of a knot
$K$ we can associate to it its Gauss diagram $G({\cal K})$. The regular
projection ${\cal K}$ can be regarded as a generic immersion of a circle
into the plane enhanced by information on the crossings. The Gauss diagram
$G({\cal K})$ consists of a circle together with the preimages of each
crossing of the immersion connected by a chord. Each chord is equipped with
the sign of the signature of the corresponding crossing. Gauss
diagrams are useful because they allow to keep track of the sums
involving the crossings which enter in (\ref{nucleos}) in a very
simple form. Let us consider a chord diagram $D$ and one of its
labeled chord subdiagrams $s\in S_D$. Let us assume that $s$ has
$p$ chords and labels
$k_1,k_2,\cdots,k_p$. We define the product,
\begin{equation}
\langle s, G({\cal K}) \rangle,
\label{producto}
\end{equation}
as the sum over all the embeddings of $s$ into $G({\cal K})$, each
one weighted by a factor,
\begin{equation}
{\varepsilon_1^{k_1} \varepsilon_2^{k_2}\cdots \varepsilon_p^{k_p} \over
(k_1! k_2! \cdots k_p!)^2},
\label{pesos}
\end{equation}
where $\varepsilon_1, \varepsilon_2, \dots, \varepsilon_p$ are the signatures
of the chords of $G({\cal K})$ involved in the embedding. Using
(\ref{producto})  the kernels $N_D(\k)$ entering (\ref{masnucleos}) can be
written as,
\begin{equation}
 N_D(\k) = \sum_{s\in S_D} m_D(s) \langle s, G({\cal K}) \rangle,
\label{laformula}
\end{equation}
where $m_D(s)$ denotes the multiplicity of the labeled subdiagram $s\in
S_D$ relative to the chord diagram $D$.

The terms $\langle s, G({\cal K}) \rangle$ entering (\ref{laformula}) are
related to the quantities $\chi({\cal K})$ defined in \cite{temporal}. It
is straightforward to obtain the following relations:
\begin{equation}
\vbox{\hskip-4cm
\vbox{\begin{eqnarray*}
\langle  \cunoj \hbox{\hskip-0.4cm}, G({\cal K}) \rangle &=& {1\over
(j!)^2}\chi_1({\cal K}),
\hbox{\hskip0.3cm} \nonumber \\
\vbox{\vskip0.9cm}
\langle \cdosii \hbox{\hskip-0.4cm}, G({\cal K}) \rangle
&=& 
\chi_2^A({\cal K}), \nonumber \\ 
\vbox{\vskip0.9cm}
\langle \cdosiidosdos \hbox{\hskip-0.4cm}, G({\cal K})
\rangle &=&  {1\over 16} \chi_2^C({\cal K}), \nonumber \\ 
\vbox{\vskip0.9cm}
\langle \ctresiii \hbox{\hskip-0.4cm}, G({\cal K})
\rangle &=& 
 \chi_3^B({\cal K}), \nonumber \\ 
\vbox{\vskip0.9cm}
\langle \ctresiiidos \hbox{\hskip-0.4cm}, G({\cal K})
\rangle &=&  {1\over 4} \chi_3^D({\cal K}), \nonumber \\ 
\vbox{\vskip0.9cm}
\langle \ccuaxi \hbox{\hskip-0.4cm}, G({\cal K}) \rangle
&=& 
\chi_4^A({\cal K}), \nonumber \\ 
\vbox{\vskip0.9cm}
\langle \ccuaix \hbox{\hskip-0.4cm}, G({\cal K}) \rangle
&=& 
\chi_4^C({\cal K}), \nonumber \\ 
\vbox{\vskip0.9cm}
\langle \ccuaviii \hbox{\hskip-0.4cm}, G({\cal K})
\rangle &=& 
\chi_4^E({\cal K}), \nonumber 
\vbox{\vskip0.9cm}
\end{eqnarray*}}

\vskip-10.3cm

\hskip2cm
\vbox{\begin{eqnarray*}
\langle \cunoj \hbox{\hskip-0.4cm}, G({\cal K}) \rangle &=& {1\over
(j!)^2}n({\cal K}),
\hbox{\hskip0.3cm}  \nonumber \\
\vbox{\vskip0.9cm}
\langle \cdosiidosuno \hbox{\hskip-0.4cm}, G({\cal K}) \rangle &=& 
{1\over 4} \chi_2^B({\cal K}), \nonumber \\ 
\vbox{\vskip0.9cm}
\langle \ctresiv \hbox{\hskip-0.4cm}, G({\cal K}) \rangle
&=& 
 \chi_3^A({\cal K}), \nonumber \\ 
\vbox{\vskip0.9cm}
\langle \ctresivdos \hbox{\hskip-0.4cm}, G({\cal K})
\rangle &=&  {1\over 4} \chi_3^C({\cal K}), \nonumber \\ 
\vbox{\vskip0.9cm}
\langle \ctresiiidosbis \hbox{\hskip-0.4cm}, G({\cal K})
\rangle &=&  {1\over 4} \chi_3^E({\cal K}), \nonumber \\ 
\vbox{\vskip0.9cm}
\langle \ccuax \hbox{\hskip-0.4cm}, G({\cal K}) \rangle
&=& 
\chi_4^B({\cal K}), \nonumber \\ 
\vbox{\vskip0.9cm}
\langle \ccuavi \hbox{\hskip-0.4cm}, G({\cal K}) \rangle
&=& 
\chi_4^D({\cal K}), \nonumber \\ 
\vbox{\vskip0.9cm}
\langle \ccuavii \hbox{\hskip-0.4cm}, G({\cal K}) \rangle
&=& 
\chi_4^F({\cal K}),  \nonumber 
\end{eqnarray*}}
}\label{lalista}
\end{equation}
where in the first row the  relation in the left column applies when $j$ is
odd and the one in the right when $j$ is even. Notice that in the second
relation
$n({\cal K})$ denotes the number of crossings of the regular projection
${\cal K}$. 

In \cite{temporal}, Vassiliev invariants up to order four were expressed in
terms of these quantities and the crossing signatures. The strategy to obtain
them was to start with the kernels (\ref{laformula}) and exploit the properties
of the perturbative series expansion of Chern-Simons gauge theory. A special
role in the construction was played by the  factorization theorem proved in
\cite{factor}. Here, only their final form will be listed.

At second order, the final expression for the invariant is:
\begin{equation}
\alpha_{21}(K) = \alpha_{21}(U) +  \langle \cdosii \hbox{\hskip-0.4cm},
G({\cal K}) \rangle - \langle \cdosii \hbox{\hskip-0.4cm},
G(\alpha({\cal K})) \rangle,
\label{primtwoca}
\end{equation} 
where  $\alpha_{21}(U)$ stands for the value of $\alpha_{21}$
for the unknot. In this expression $\alpha({\k})$ denotes the
ascending diagram
$\alpha({\k})$ of a knot projection $\k$. It is defined as the
diagram obtained by switching, when traveling along the knot
from a base point,  all the undercrossings to overcrossings.
Ascending diagrams enter often in the combinatorial expressions
and it is convenient introduce the following notation. A bar
over a quantity
$L(\k)$ indicates that the same quantity for the ascending
diagram has to be subtracted, \ie:
\begin{equation}
\bar L(\k) = L(\k) - L(\alpha(\k))
\label{limon}
\end{equation}
where $\alpha({\cal K})$ denotes the standard ascending diagram
of ${\cal K}$. Using this notation, the final form for the only
primitive Vassiliev invariant at order two is:
\begin{equation}
\alpha_{21}(K) = \alpha_{21}(U) +  
\langle \cdosii \hbox{\hskip-0.4cm}, \bar
G(\k) \rangle.
\label{primtwoc}
\end{equation}

 At order three there is only one
primitive invariant. It takes the form:
\begin{equation}
\alpha_{31} (K) = \langle \ctresiii \hbox{\hskip-0.4cm} + \ctresiv
\hbox{\hskip-0.4cm} + 2 \cdosiidosuno
\hbox{\hskip-0.4cm},  G(\k) \rangle
 - \sum_{i=1}^n \, \varepsilon_i(\k) \Big[ \langle  \cdosii
\hbox{\hskip-0.4cm}, G(\alpha(\k)) \rangle \Big]_i.
\label{primthreeb}
\end{equation}
Several comments are in order to explain the quantities entering  this
expression. The sum is
over all crossings $i$, $i=1,\dots,n$, and $\varepsilon_i(\k)$ denotes the
corresponding signature. The square brackets $[$ $]_i$ enclosing a quantity
$L(\k)$ denote:
\begin{equation}
 \Big[  L(\k) \Big]_i = L(\k) - L(\k_{i_+}) -L(\k_{i_-}),
\label{platano}
\end{equation}
where the regular projection diagrams $\k _{i_+}$ and
$\k _{i_-}$ are the ones which result after the splitting of $\k$ at the
crossing point $i$. 

Combinatorial expressions for the two primitive invariants at
order four have been presented in \cite{temporal}. Their
construction is based on the use of the kernels
(\ref{laformula}) and the properties of the perturbative series
expansion. As in the case of previous orders, these invariants
are expressed in terms of the products (\ref{producto}) and the
crossing signatures. Their form is more complicated than the
ones at lower orders.
 At order four there are two primitive Vassiliev
invariants. The same choice of basis as in
\cite{temporal} is made here.  The combinatorial expressions for these two
invariants turn out to be:
\begin{eqnarray}
&& {\hskip -0.5cm} \alpha_{42}(K) = \alpha_{42}(U) +
\langle 7 \ccuaxi \hbox{\hskip-0.4cm} + 5 \ccuax \hbox{\hskip-0.4cm} + 4
\ccuaix \hbox{\hskip-0.4cm} + 2 \ccuaviii \hbox{\hskip-0.4cm} + \ccuavi
\hbox{\hskip-0.4cm} + \ccuavii \hbox{\hskip-0.4cm}  \nonumber \\ && 
\nonumber \\ && 
\hbox{\hskip4.2cm} + 8
\ctresivdos
\hbox{\hskip-0.4cm} + 2 \ctresiiidos \hbox{\hskip-0.4cm} + 8 \ctresiiidosbis
\hbox{\hskip-0.4cm} +{1\over 6} \cdosii \hbox{\hskip-0.4cm},
\bar G(\k) \rangle  \nonumber \\ && {\hskip -0.5cm} +
\sum_{i,j\in {\cal C}_a\atop i>j} \bar\varepsilon_{ij}(\k)\Bigg(
\Big[\langle \cdosii \hbox{\hskip-0.4cm}, G(\alpha(\k))\rangle \Big]_{ij}^a
\nonumber \\ &&
\nonumber \\ &&
\hbox{\hskip3cm} -2
\Big[\langle \cdosii \hbox{\hskip-0.4cm}, G(\alpha(\k))\rangle\Big]_i - 2
\Big[\langle \cdosii \hbox{\hskip-0.4cm}, G(\alpha(\k))\rangle\Big]_j
\Bigg)
\nonumber \\ && {\hskip -0.5cm} +
\sum_{i,j\in {\cal C}_b\atop i>j} \bar\varepsilon_{ij}(\k)\Bigg(
\Big[\langle \cdosii \hbox{\hskip-0.4cm}, G(\alpha(\k))\rangle\Big]_{ij}^b -
\Big[\langle \cdosii \hbox{\hskip-0.4cm}, G(\alpha(\k))\rangle\Big]_i 
\nonumber \\ &&
\nonumber \\ &&
\hbox{\hskip7.3cm} - 
\Big[\langle \cdosii \hbox{\hskip-0.4cm}, G(\alpha(\k))\rangle\Big]_j
\Bigg),  \nonumber \\ 
\label{primfourm}
\end{eqnarray}
and,
\begin{eqnarray}
&& {\hskip -0.1cm} \alpha_{43}(K) = \alpha_{43}(U) + 
\langle  \ccuaxi \hbox{\hskip-0.4cm} + \ccuax \hbox{\hskip-0.4cm} + 
\ccuaix \hbox{\hskip-0.4cm} + 2 \ctresiiidosbis \hbox{\hskip-0.4cm}
- {1\over 6} \cdosii \hbox{\hskip-0.4cm}, \bar G(\k) \rangle
\nonumber \\ && +
\sum_{i,j\in {\cal C}_a\atop i>j} \bar\varepsilon_{ij}(\k)\Bigg(
\Big[\langle \cdosii \hbox{\hskip-0.4cm},
G(\alpha(\k))\rangle\Big]_{ij}^a-\Big[\langle \cdosii \hbox{\hskip-0.4cm},
G(\alpha(\k))\rangle\Big]_i 
\nonumber \\ &&
\nonumber \\ &&
\hbox{\hskip7cm} - 
\Big[\langle \cdosii \hbox{\hskip-0.4cm}, G(\alpha(\k))\rangle\Big]_j
\Bigg). \nonumber \\
\label{primfourn}
\end{eqnarray}
In these expressions the explicit dependence on the signatures appears in
the quantities $\bar\varepsilon_{ij}(\k)$ which are:
\begin{equation}
\bar\varepsilon_{ij}(\k) = \varepsilon_{ij}(\k) - \varepsilon_{ij}(\alpha(\k))=
\varepsilon_{i}(\k)\varepsilon_{j}(\k) -
\varepsilon_{i}(\alpha(\k))\varepsilon_{j}(\alpha(\k)).
\label{lasepsilons}
\end{equation}
The sums in which these products are involved are over double splittings of
the knot projection $\k$ at the crossings $i$ and $j$. There are two ways of
carrying out these double splittings, depending on the configuration
associated to the crossings $i$ and $j$. These are described in detail in
\cite{temporal}. In the first one the regular projection $\k$ is split  into
two while in the second one it is split into three. Splittings of the first
type build the set
${\cal C}_a$. The ones of the second type build ${\cal C}_b$. While only
the first one is involved in the invariant $\alpha_{43}$, both appear
in $\alpha_{42}$. The new quantities entering the sums are:
\begin{eqnarray}
\Big[L(\k)\Big]_{ij}^a &=&
L(\k) - L(\k_{ij}^{a_1}) -  L(\k_{ij}^{a_2}),
\nonumber \\
\Big[L(\k)\Big]_{ij}^b &=&
L(\k) - L(\k_{ij}^{b_1}) -  L(\k_{ij}^{b_2}) -  L(\k_{ij}^{b_3}),
\label{sandia}
\end{eqnarray}
where $\k_{ij}^{a_1},\k_{ij}^{a_2},\k_{ij}^{b_1},\k_{ij}^{b_2}$ and
$\k_{ij}^{b_3}$ are the knot projections which originate after a double
splitting of $\k$. As in previous orders, in the  expressions
(\ref{primfourm}) and (\ref{primfourn}), the quantities
$\alpha_{42}(U)$ and
$\alpha_{43}(U)$ correspond to the value of these invariants for the
unknot. It has been proved in \cite{temporal} that the
combinatorial expressions for $\alpha_{42}(K)$ and
$\alpha_{43}(K)$ in (\ref{primfourm}) and (\ref{primfourn}) are
invariant under Reidemeister moves.

Vassiliev invariants constitute vector spaces and their normalization can
be chosen in such a way that they are integer-valued. Once their value for
the unknot has been subtracted off they can be presented in many
basis in which they are integers. We will chose here a  particular basis in
which the numerical values for the invariants up to order four are rather
simple:
\begin{eqnarray}
\nu_{2}(K) &=& {1\over 4} \tilde\alpha_{21}(K), \nonumber \\
\nu_{3}(K) &=& {1\over 8} \tilde\alpha_{31}(K), \nonumber \\
\nu_{4}^1(K) &=& {1\over 8} (\tilde\alpha_{42}(K) + \tilde\alpha_{43}(K)),
\nonumber \\
\nu_{4}^2(K) &=& {1\over 4} (\tilde\alpha_{42}(K) - 5
\tilde\alpha_{43}(K)).
\label{basica}
\end{eqnarray}
In these equations the tilde indicates that the value for the unknot
has been subtracted, \ie,
$\tilde\alpha_{ij}(K)=\alpha_{ij}(K)-\alpha_{ij}(U)$. The values of the
Vassiliev invariants (\ref{basica}) for all prime knots up to nine crossings
have been presented in \cite{baires}.  The invariants (\ref{basica}) have
been computed for torus knots in
\cite{torusknots} and \cite{simon}. Denoting a generic torus knot by two
coprime integers, $p$ and $q$, these invariants take the form:
\begin{eqnarray}
\nu_2(p,q) &=& {1\over 24}  (p^2-1)(q^2-1), \nonumber \\
\nu_3(p,q) &=&  {1\over 144} (p^2-1)(q^2-1)pq, \nonumber \\
\nu_4^1(p,q) &=& {1\over 288} (p^2-1)(q^2-1)p^2q^2, \\
\nu_4^2(p,q) &=& {1\over 720} (p^2-1)(q^2-1)(2p^2q^2-3p^2-3q^2-3). \nonumber
\label{toros}
\end{eqnarray}
The explicit expression of Vassiliev invariants as polynomials in $p$ and
$q$ is known up to order six \cite{torusknots}. Of course, up to order four
their value agree with the ones computed explicitly from equations
(\ref{primtwoc}), (\ref{primthreeb}), (\ref{primfourm}) and
(\ref{primfourn}). The only torus knots up to ten crossings are $3_1$,
$5_1$, $7_1$, $8_{19}$, $9_1$ and $10_{124}$, whose associated coprime
integers are (3,2), (5,2), (7,2), (4,3), (9,2) and (5,3), respectively.

In the table collected in the appendix the value of the
primitive Vassiliev invariants for all the prime knots with ten
crossings are presented. The value of these invariants for
prime knots up to ten crossings can be found in \cite{baires}.

\section{Prospects}


Though the perspectives are rather promising, the problems inherent to the
proper treatment of gauge theories in non-covariant gauges do not permit at
the moment to obtain a closed and complete formulation. Much work has to be
done to understand the subtle issues related to the use of non-covariant
gauges. The kernels plus the properties of the perturbative series expansion
are probably enough to compute the explicit form of a given invariant but
certainly it does not provide a systematic way of deriving the general
universal formula. A proper and complete formulation of the perturbative
series in a non-covariant gauge would explain the presence of the Kontsevich
factor and will provide a general universal combinatorial formula. 
It is likely that a lattice formulation of Chern-Simons gauge theory in the
temporal gauge could help considerably to make progress in this direction. We
expect to report on this and other issues of perturbative Chern-Simons gauge
theory in future work.


{\section*{Acknowledgments}}
\noindent 
I would like to thank the organizers of the Workshop
on Geometry and Physics for inviting me to deliver
a lecture and for their warm hospitality. This work is
supported in part by DGICYT under grant PB96-0960.

{\section*{Appendix}}

In this appendix the result of computing the first four primitive Vassiliev
invariants in (\ref{basica}), after using the expressions 
(\ref{primtwoc}), (\ref{primthreeb}), (\ref{primfourm}) and
(\ref{primfourn}), for all prime knots with 10 crossings, is presented.
These quantities have not been computed before using these combinatorial
expressions. Their calculation is lengthy but straightforward once the
computation of (\ref{primtwoc}), (\ref{primthreeb}), (\ref{primfourm}) and
(\ref{primfourn}) are programmed and the
prime knots are properly labeled.

\begin{table}[hp]
\begin{center}
\begin{tabular}{|c||c|c|c|c|c|c||c|c|c|c|}\cline{1-5} \cline{7-11}
  Knot & $\nu_2$ & $\nu_3$ & $\nu_4^1$ & $\nu_4^2$  & $\;\;\;\;\;\;$
& Knot & $\nu_2$ & $\nu_3$ & $\nu_4^1$ & $\nu_4^2$ \\
\cline{1-5} \cline{7-11}	$	10_{1	}	$	&	$-$4	&	$-$6	&	18	&	$-$64	&&	$	10_{41	}	$	&	$-$2	&	$-$2	&	6	&	$-$26	\\
\cline{1-5} \cline{7-11}	$	10_{2	}	$	&	2	&	2	&	18	&	$-$46	&&	$	10_{42	}	$	&	0	&	$-$1	&	$-$1	&	10	\\
\cline{1-5} \cline{7-11}	$	10_{3	}	$	&	$-$6	&	$-$3	&	65	&	$-$92	&&	$	10_{43	}	$	&	2	&	0	&	8	&	6	\\
\cline{1-5} \cline{7-11}	$	10_{4	}	$	&	$-$5	&	$-$1	&	53	&	$-$85	&&	$	10_{44	}	$	&	0	&	1	&	3	&	$-$6	\\
\cline{1-5} \cline{7-11}	$	10_{5	}	$	&	4	&	$-$7	&	31	&	54	&&	$	10_{45	}	$	&	$-$2	&	0	&	6	&	$-$2	\\
\cline{1-5} \cline{7-11}	$	10_{6	}	$	&	$-$1	&	$-$4	&	2	&	$-$75	&&	$	10_{46	}	$	&	0	&	$-$4	&	$-$6	&	$-$84	\\
\cline{1-5} \cline{7-11}	$	10_{7	}	$	&	$-$1	&	$-$3	&	$-$1	&	$-$45	&&	$	10_{47	}	$	&	6	&	$-$11	&	77	&	40	\\
\cline{1-5} \cline{7-11}	$	10_{8	}	$	&	$-$3	&	$-$4	&	26	&	$-$89	&&	$	10_{48	}	$	&	4	&	0	&	16	&	36	\\
\cline{1-5} \cline{7-11}	$	10_{9	}	$	&	$-$2	&	$-$2	&	20	&	$-$70	&&	$	10_{49	}	$	&	7	&	16	&	112	&	57	\\
\cline{1-5} \cline{7-11}	$	10_{10	}	$	&	1	&	$-$2	&	0	&	31	&&	$	10_{50	}	$	&	$-$1	&	$-$5	&	$-$5	&	$-$77	\\
\cline{1-5} \cline{7-11}	$	10_{11	}	$	&	$-$5	&	$-$4	&	52	&	$-$99	&&	$	10_{51	}	$	&	5	&	$-$8	&	54	&	23	\\
\cline{1-5} \cline{7-11}	$	10_{12	}	$	&	4	&	$-$6	&	30	&	40	&&	$	10_{52	}	$	&	3	&	1	&	9	&	27	\\
\cline{1-5} \cline{7-11}	$	10_{13	}	$	&	$-$5	&	$-$2	&	40	&	$-$51	&&	$	10_{53	}	$	&	6	&	13	&	87	&	36	\\
\cline{1-5} \cline{7-11}	$	10_{14	}	$	&	2	&	3	&	21	&	$-$28	&&	$	10_{54	}	$	&	4	&	$-$2	&	22	&	24	\\
\cline{1-5} \cline{7-11}	$	10_{15	}	$	&	3	&	$-$2	&	8	&	37	&&	$	10_{55	}	$	&	5	&	10	&	58	&	31	\\
\cline{1-5} \cline{7-11}	$	10_{16	}	$	&	$-$4	&	$-$4	&	34	&	$-$80	&&	$	10_{56	}	$	&	0	&	$-$2	&	0	&	$-$48	\\
\cline{1-5} \cline{7-11}	$	10_{17	}	$	&	2	&	0	&	$-$6	&	50	&&	$	10_{57	}	$	&	4	&	$-$6	&	34	&	24	\\
\cline{1-5} \cline{7-11}	$	10_{18	}	$	&	$-$2	&	$-$1	&	15	&	$-$44	&&	$	10_{58	}	$	&	$-$4	&	$-$1	&	21	&	$-$22	\\
\cline{1-5} \cline{7-11}	$	10_{19	}	$	&	1	&	0	&	$-$8	&	39	&&	$	10_{59	}	$	&	$-$1	&	$-$1	&	1	&	$-$17	\\
\cline{1-5} \cline{7-11}	$	10_{20	}	$	&	$-$3	&	$-$6	&	10	&	$-$73	&&	$	10_{60	}	$	&	$-$1	&	1	&	$-$3	&	$-$1	\\
\cline{1-5} \cline{7-11}	$	10_{21	}	$	&	1	&	0	&	8	&	$-$49	&&	$	10_{61	}	$	&	$-$4	&	$-$5	&	39	&	$-$106	\\
\cline{1-5} \cline{7-11}	$	10_{22	}	$	&	$-$4	&	$-$2	&	42	&	$-$88	&&	$	10_{62	}	$	&	5	&	$-$9	&	51	&	53	\\
\cline{1-5} \cline{7-11}	$	10_{23	}	$	&	3	&	$-$5	&	17	&	43	&&	$	10_{63	}	$	&	6	&	14	&	96	&	42	\\
\cline{1-5} \cline{7-11}	$	10_{24	}	$	&	$-$2	&	$-$5	&	3	&	$-$68	&&	$	10_{64	}	$	&	$-$3	&	$-$3	&	31	&	$-$91	\\
\cline{1-5} \cline{7-11}	$	10_{25	}	$	&	0	&	$-$2	&	0	&	$-$48	&&	$	10_{65	}	$	&	4	&	$-$7	&	35	&	38	\\
\cline{1-5} \cline{7-11}	$	10_{26	}	$	&	$-$3	&	$-$2	&	26	&	$-$65	&&	$	10_{66	}	$	&	7	&	17	&	123	&	67	\\
\cline{1-5} \cline{7-11}	$	10_{27	}	$	&	2	&	$-$3	&	5	&	36	&&	$	10_{67	}	$	&	0	&	0	&	8	&	$-$32	\\
\cline{1-5} \cline{7-11}	$	10_{28	}	$	&	3	&	$-$4	&	16	&	29	&&	$	10_{68	}	$	&	2	&	$-$3	&	5	&	36	\\
\cline{1-5} \cline{7-11}	$	10_{29	}	$	&	$-$4	&	$-$3	&	29	&	$-$54	&&	$	10_{69	}	$	&	2	&	$-$4	&	20	&	6	\\
\cline{1-5} \cline{7-11}	$	10_{30	}	$	&	1	&	1	&	11	&	$-$31	&&	$	10_{70	}	$	&	$-$3	&	2	&	16	&	$-$37	\\
\cline{1-5} \cline{7-11}	$	10_{31	}	$	&	2	&	$-$1	&	1	&	28	&&	$	10_{71	}	$	&	1	&	0	&	2	&	11	\\
\cline{1-5} \cline{7-11}	$	10_{32	}	$	&	$-$1	&	0	&	10	&	$-$35	&&	$	10_{72	}	$	&	2	&	4	&	24	&	$-$10	\\
\cline{1-5} \cline{7-11}	$	10_{33	}	$	&	0	&	0	&	$-$8	&	32	&&	$	10_{73	}	$	&	1	&	$-$2	&	10	&	3	\\
\cline{1-5} \cline{7-11}	$	10_{34	}	$	&	3	&	$-$3	&	17	&	19	&&	$	10_{74	}	$	&	0	&	$-$2	&	0	&	$-$48	\\
\cline{1-5} \cline{7-11}	$	10_{35	}	$	&	$-$4	&	2	&	22	&	$-$32	&&	$	10_{75	}	$	&	0	&	1	&	$-$5	&	2	\\
\cline{1-5} \cline{7-11}	$	10_{36	}	$	&	1	&	2	&	12	&	$-$17	&&	$	10_{76	}	$	&	$-$2	&	$-$6	&	2	&	$-$82	\\
\cline{1-5} \cline{7-11}	$	10_{37	}	$	&	3	&	0	&	12	&	21	&&	$	10_{77	}	$	&	4	&	$-$5	&	29	&	26	\\
\cline{1-5} \cline{7-11}	$	10_{38	}	$	&	$-$1	&	$-$2	&	6	&	$-$43	&&	$	10_{78	}	$	&	3	&	5	&	23	&	7	\\
\cline{1-5} \cline{7-11}	$	10_{39	}	$	&	1	&	1	&	11	&	$-$31	&&	$	10_{79	}	$	&	5	&	0	&	32	&	27	\\
\cline{1-5} \cline{7-11}	$	10_{40	}	$	&	3	&	$-$4	&	16	&	29	&&	$	10_{80	}	$	&	6	&	12	&	72	&	42	\\
\cline{1-5} \cline{7-11}
\end{tabular}
\caption{Primitive Vassiliev invariants up to order four
for all prime knots with ten crossings.}
\end{center}
\label{tablauno}
\end{table}

\begin{table}[hp]
\begin{center}
\begin{tabular}{|c||c|c|c|c|c|c||c|c|c|c|}\cline{1-5} \cline{7-11}
  Knot & $\nu_2$ & $\nu_3$ & $\nu_4^1$ & $\nu_4^2$  & $\;\;\;\;\;\;$
& Knot & $\nu_2$ & $\nu_3$ & $\nu_4^1$ & $\nu_4^2$ \\
\cline{1-5} \cline{7-11}	$	10_{81	}	$	&	3	&	0	&	16	&	5	&&	$	10_{124	}	$	&	8	&	20	&	150	&	92	\\
\cline{1-5} \cline{7-11}	$	10_{82	}	$	&	0	&	0	&	8	&	$-$32	&&	$	10_{125	}	$	&	3	&	0	&	10	&	17	\\
\cline{1-5} \cline{7-11}	$	10_{83	}	$	&	1	&	$-$2	&	0	&	31	&&	$	10_{126	}	$	&	5	&	$-$9	&	59	&	21	\\
\cline{1-5} \cline{7-11}	$	10_{84	}	$	&	2	&	2	&	4	&	22	&&	$	10_{127	}	$	&	1	&	$-$1	&	$-$3	&	$-$35	\\
\cline{1-5} \cline{7-11}	$	10_{85	}	$	&	2	&	$-$3	&	5	&	36	&&	$	10_{128	}	$	&	7	&	17	&	123	&	67	\\
\cline{1-5} \cline{7-11}	$	10_{86	}	$	&	$-$1	&	$-$1	&	7	&	$-$29	&&	$	10_{129	}	$	&	2	&	1	&	5	&	12	\\
\cline{1-5} \cline{7-11}	$	10_{87	}	$	&	0	&	1	&	7	&	$-$22	&&	$	10_{130	}	$	&	4	&	$-$6	&	38	&	8	\\
\cline{1-5} \cline{7-11}	$	10_{88	}	$	&	$-$1	&	0	&	$-$2	&	13	&&	$	10_{131	}	$	&	0	&	$-$2	&	$-$4	&	$-$32	\\
\cline{1-5} \cline{7-11}	$	10_{89	}	$	&	1	&	$-$3	&	15	&	1	&&	$	10_{132	}	$	&	3	&	$-$5	&	25	&	11	\\
\cline{1-5} \cline{7-11}	$	10_{90	}	$	&	$-$3	&	$-$1	&	25	&	$-$55	&&	$	10_{133	}	$	&	1	&	0	&	0	&	$-$17	\\
\cline{1-5} \cline{7-11}	$	10_{91	}	$	&	2	&	0	&	$-$2	&	34	&&	$	10_{134	}	$	&	6	&	13	&	83	&	52	\\
\cline{1-5} \cline{7-11}	$	10_{92	}	$	&	2	&	3	&	17	&	$-$12	&&	$	10_{135	}	$	&	3	&	1	&	13	&	11	\\
\cline{1-5} \cline{7-11}	$	10_{93	}	$	&	1	&	$-$1	&	$-$5	&	33	&&	$	10_{136	}	$	&	0	&	$-$1	&	3	&	$-$6	\\
\cline{1-5} \cline{7-11}	$	10_{94	}	$	&	$-$2	&	$-$2	&	16	&	$-$54	&&	$	10_{137	}	$	&	$-$2	&	$-$2	&	2	&	$-$10	\\
\cline{1-5} \cline{7-11}	$	10_{95	}	$	&	3	&	$-$5	&	21	&	27	&&	$	10_{138	}	$	&	$-$3	&	$-$2	&	12	&	$-$21	\\
\cline{1-5} \cline{7-11}	$	10_{96	}	$	&	$-$3	&	2	&	12	&	$-$21	&&	$	10_{139	}	$	&	9	&	25	&	209	&	109	\\
\cline{1-5} \cline{7-11}	$	10_{97	}	$	&	2	&	4	&	28	&	$-$26	&&	$	10_{140	}	$	&	2	&	4	&	14	&	18	\\
\cline{1-5} \cline{7-11}	$	10_{98	}	$	&	0	&	$-$3	&	$-$7	&	$-$50	&&	$	10_{141	}	$	&	$-$1	&	$-$1	&	7	&	$-$29	\\
\cline{1-5} \cline{7-11}	$	10_{99	}	$	&	4	&	0	&	20	&	20	&&	$	10_{142	}	$	&	8	&	21	&	169	&	70	\\
\cline{1-5} \cline{7-11}	$	10_{10	}	$	&	4	&	$-$7	&	35	&	38	&&	$	10_{143	}	$	&	3	&	$-$5	&	21	&	27	\\
\cline{1-5} \cline{7-11}	$	10_{101	}	$	&	7	&	17	&	127	&	51	&&	$	10_{144	}	$	&	$-$2	&	$-$2	&	12	&	$-$38	\\
\cline{1-5} \cline{7-11}	$	10_{102	}	$	&	$-$2	&	$-$1	&	15	&	$-$44	&&	$	10_{145	}	$	&	5	&	$-$12	&	82	&	31	\\
\cline{1-5} \cline{7-11}	$	10_{103	}	$	&	3	&	$-$4	&	16	&	29	&&	$	10_{146	}	$	&	0	&	0	&	$-$4	&	16	\\
\cline{1-5} \cline{7-11}	$	10_{104	}	$	&	1	&	0	&	$-$8	&	39	&&	$	10_{147	}	$	&	$-$1	&	0	&	6	&	$-$19	\\
\cline{1-5} \cline{7-11}	$	10_{105	}	$	&	$-$1	&	0	&	6	&	$-$19	&&	$	10_{148	}	$	&	4	&	$-$7	&	39	&	22	\\
\cline{1-5} \cline{7-11}	$	10_{106	}	$	&	$-$1	&	$-$1	&	11	&	$-$45	&&	$	10_{149	}	$	&	2	&	2	&	10	&	$-$14	\\
\cline{1-5} \cline{7-11}	$	10_{107	}	$	&	1	&	$-$1	&	$-$1	&	17	&&	$	10_{150	}	$	&	1	&	1	&	7	&	$-$15	\\
\cline{1-5} \cline{7-11}	$	10_{108	}	$	&	0	&	0	&	$-$8	&	32	&&	$	10_{151	}	$	&	3	&	$-$4	&	20	&	13	\\
\cline{1-5} \cline{7-11}	$	10_{109	}	$	&	3	&	0	&	6	&	33	&&	$	10_{152	}	$	&	7	&	15	&	97	&	63	\\
\cline{1-5} \cline{7-11}	$	10_{110	}	$	&	$-$3	&	$-$3	&	17	&	$-$47	&&	$	10_{153	}	$	&	4	&	$-$1	&	23	&	14	\\
\cline{1-5} \cline{7-11}	$	10_{111	}	$	&	1	&	0	&	4	&	$-$33	&&	$	10_{154	}	$	&	5	&	9	&	49	&	25	\\
\cline{1-5} \cline{7-11}	$	10_{112	}	$	&	2	&	$-$2	&	12	&	$-$10	&&	$	10_{155	}	$	&	$-$2	&	2	&	12	&	$-$38	\\
\cline{1-5} \cline{7-11}	$	10_{113	}	$	&	0	&	$-$1	&	$-$5	&	2	&&	$	10_{156	}	$	&	1	&	$-$1	&	$-$1	&	17	\\
\cline{1-5} \cline{7-11}	$	10_{114	}	$	&	1	&	$-$1	&	7	&	$-$15	&&	$	10_{157	}	$	&	4	&	$-$8	&	50	&	8	\\
\cline{1-5} \cline{7-11}	$	10_{115	}	$	&	1	&	0	&	$-$4	&	23	&&	$	10_{158	}	$	&	$-$3	&	$-$1	&	21	&	$-$39	\\
\cline{1-5} \cline{7-11}	$	10_{116	}	$	&	0	&	0	&	4	&	$-$16	&&	$	10_{159	}	$	&	2	&	3	&	9	&	20	\\
\cline{1-5} \cline{7-11}	$	10_{117	}	$	&	2	&	$-$3	&	9	&	20	&&	$	10_{160	}	$	&	3	&	6	&	36	&	$-$3	\\
\cline{1-5} \cline{7-11}	$	10_{118	}	$	&	0	&	0	&	$-$4	&	16	&&	$	10_{161	}	$	&	7	&	18	&	138	&	61	\\
\cline{1-5} \cline{7-11}	$	10_{119	}	$	&	$-$1	&	0	&	6	&	$-$19	&&	$	10_{162	}	$	&	$-$3	&	4	&	18	&	$-$57	\\
\cline{1-5} \cline{7-11}	$	10_{120	}	$	&	6	&	13	&	83	&	52	&&	$	10_{163	}	$	&	1	&	2	&	4	&	15	\\
\cline{1-5} \cline{7-11}	$	10_{121	}	$	&	1	&	2	&	4	&	15	&&	$	10_{164	}	$	&	1	&	0	&	$-$4	&	23	\\
\cline{1-5} \cline{7-11}	$	10_{122	}	$	&	2	&	$-$2	&	12	&	$-$10	&&	$	10_{165	}	$	&	2	&	$-$3	&	17	&	$-$12	\\
\cline{1-5} \cline{7-11}	$	10_{123	}	$	&	$-$2	&	0	&	10	&	$-$18	&&	$			$	&		&		&		&		\\
\cline{1-5} \cline{7-11}
\end{tabular}
\caption{Primitive Vassiliev invariants up to order four
for all prime knots with ten crossings.}
\end{center}
\label{tablauno}
\end{table}

%
%


\end{document}